\documentclass[aps,prd,showpacs,onecolumn,preprintnumbers,notitlepage,
nofootinbib,tightenlines,10pt]{revtex4-1}
\usepackage{amsmath,bm}
\usepackage{times}
\usepackage{amsfonts}
\usepackage{amssymb}
\usepackage{braket}
\usepackage{color,graphicx}
\usepackage{epstopdf}
\usepackage{latexsym}
\usepackage[dvipsnames]{xcolor}
\usepackage{slashed}
\usepackage{CJK,upgreek,fancyhdr}
\usepackage{float}
\usepackage{multirow}
\usepackage{hyperref}
\hypersetup{colorlinks,citecolor=nicegreen,linkcolor=nicered,urlcolor=RoyalBlue}
\usepackage{enumitem}
\usepackage{natbib}
\usepackage{relsize}
\usepackage[left=2.5cm,right=2.5cm,top=2.0cm,bottom=2.5cm]{geometry}
\newcommand{\beq}{\begin{eqnarray}}
\newcommand{\eeq}{\end{eqnarray}}
\newcommand{\non}{\nonumber\\ }

\newcommand{\psl}{ P \hspace{-2.6truemm}/ }

\def\lsim{ {\ \lower-1.2pt\vbox{\hbox{\rlap{$<$}\lower6pt\vbox{\hbox{$\sim$}
}}}\ } }
\def\gsim{ {\ \lower-1.2pt\vbox{\hbox{\rlap{$>$}\lower6pt\vbox{\hbox{$\sim$}
}}}\ } }

\def \jhep{ J. High Energy Phys.  }

\definecolor{Red}{rgb}{1.,0.,0.}
\definecolor{Blue}{rgb}{0.,0.,1.}
\definecolor{RoyalBlue}{rgb}{0.0, 0.14, 0.4}
\definecolor{nicered}{rgb}{0.7,0.1,0.2}
\definecolor{nicegreen}{rgb}{0.1,0.4,0.2}

\newcommand{\Gre}[1]{{\color{nicegreen}{#1}}}

\def\orcid#1{\kern .08em\href{https://orcid.org/#1}{\includegraphics[keepaspectratio,width=0.75em]{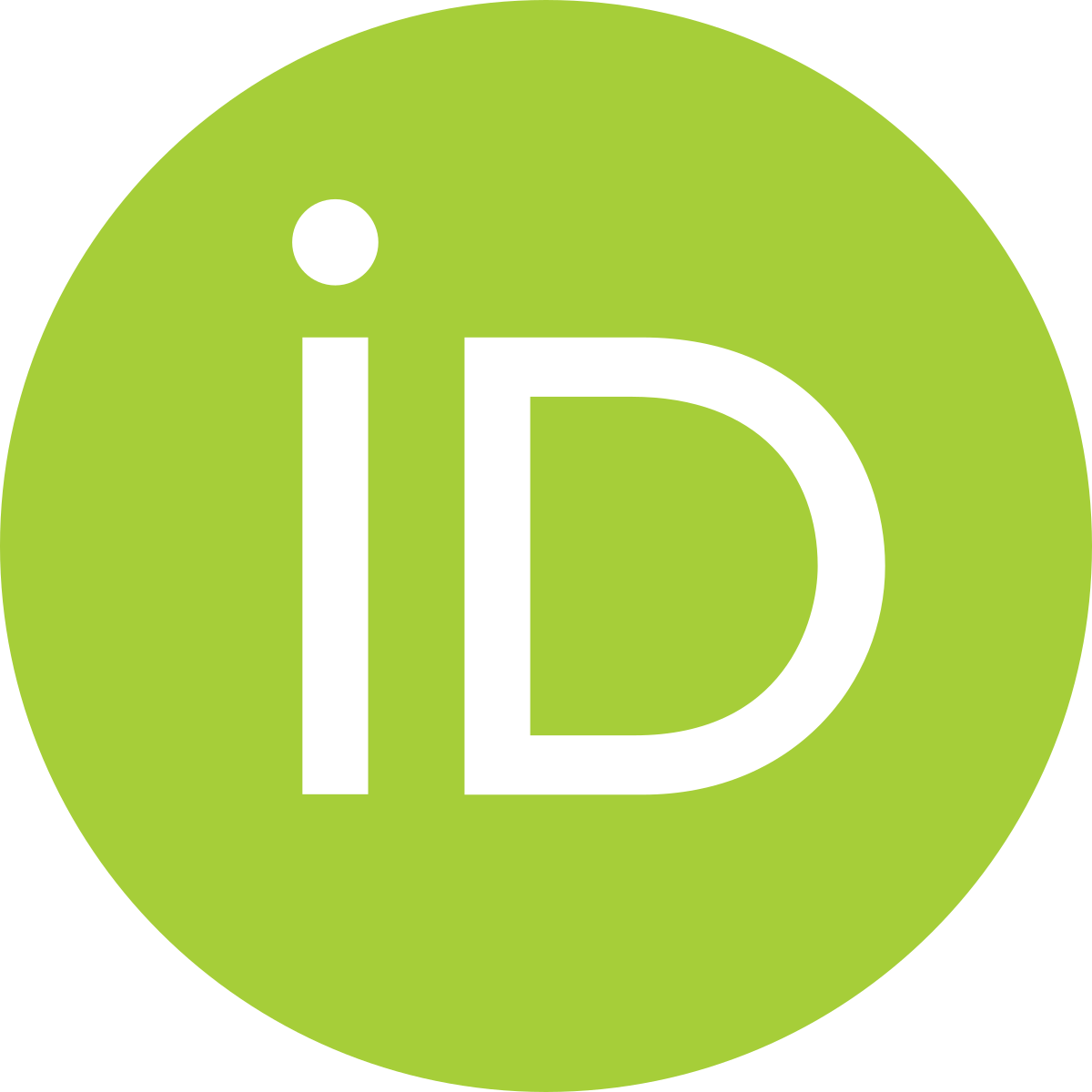}}}

\begin{document}
\begin{CJK*}{GB}{gbsn}
\title{\boldmath Improved perturbative QCD study of the decay $B_c^+ \to \eta_c L^+$}
\author{Wen-Jing~Zhang\orcid{0009-0000-1210-8420}}
\author{Xin~Liu\orcid{0000-0001-9419-7462}}
\affiliation{Department of Physics,
Jiangsu Normal University, Xuzhou 221116, China}


\date{\today{}}

\begin{abstract}

We perform an improved perturbative QCD study of the decays $B_c^+ \to \eta_c L^+$,
where $L$ denotes the light ground-state pseudoscalar, vector mesons and the corresponding $p$-wave
scalar, axial-vector, and tensor ones, and predict their branching ratios (BRs) associated with
relative ratios at leading order in the strong coupling $\alpha_s$. Our results ${\rm BR}
(B_c^+ \to \eta_c \pi^+) =(2.03^{+0.53}_{-0.41}) \times 10^{-3}$ and ${\rm BR}(B_c^+ \to \eta_c
\pi^+)/{\rm BR}(B_c^+ \to J/\psi \pi^+) = 1.74^{+0.66}_{-0.50}$ are consistent with several
available predictions in different approaches within uncertainties. Inputting the measured $\eta_c
\to p\bar p$ and $\eta_c \to \pi^+\pi^- (\pi^+\pi^-, K^+ K^-, p\bar p)$ BRs with $p$ here being
a proton, we derive the multibody $B_c^+ \to \eta_c (\pi, \rho)^+$ BRs through secondary decay
chains via resonance $\eta_c$ under the narrow-width approximation, which might facilitate
the (near-) future tests of $B_c \to \eta_c$ decays. Under the $q\bar q$ assignment
for light scalars, different to $B_c$ decaying into $J/\psi$ plus a scalar meson and other
$B_c^+ \to \eta_c L^+$ modes, surprisingly small $\Delta S =0$ BRs around ${\cal O}(10^{-7}-10^{-9})$
and highly large ratios near ${\cal O}(10^{2})$ between the $\Delta S=1$ and $\Delta S=0$ BRs
are found in the $B_c$ decays to $\eta_c$ plus light scalars, with $S$ being strange number.
Many large BRs and interesting ratios presented in this work could be tested by
the Large Hadron Collider experiments, which would help us to examine
the reliability of this improved perturbative QCD
formalism for $B_c$-meson decays and further understand the QCD dynamics in the considered
decay modes as well as in the related hadrons.

\end{abstract}


\pacs{13.25.Hw, 12.38.Bx, 14.40.Nd}
\preprint{\footnotesize  JSNU-PHY-HEP-OCT.25}
\maketitle


\newpage
%
%

\section{Introduction}\label{sec:Intro}

It is well known that, disparate from the $b\bar b$ and $c\bar c$ states, the $B_c$ meson
is flavor antisymmetric while unique since it is the only ground state containing two different
heavy quarks $b$ and $c$ simultaneously~\cite{Brambilla:2004wf,Brambilla:2010cs}. Peculiar to
the more extensively studied $B_u$, $B_d$, and $B_s$ mesons, both constituents in a
$B_c$ meson can decay separately, which offers a precious
opportunity to expand our understanding
of heavy $B$-meson physics through thoroughly exploring the more rich and complicated
QCD in the perturbative and nonperturbative regimes. Thus, the weak decays
of the $B_c$ meson has prompted significant theoretical and experimental interest since its
first discovery at the Tevatron in 1998~\cite{CDF:1998axz,CDF:1998ihx}.

On the experimental side, the LHC experiments that started running in
2009 have observed many $B_c$-meson decay channels of interest, such as $B_c^+ \to J/\psi
(\pi, 2\pi, 3\pi)^+$, even $B_c^+ \to \chi_{cJ} \pi^+ (J=0,1,2)$, etc.~\cite{ParticleDataGroup:2024cfk,HeavyFlavorAveragingGroupHFLAV:2024ctg}. The $\eta_c$
meson, the lowest-lying $c\bar c$ pseudoscalar state, has attracted considerable theoretical
and experimental attention since its discovery~\cite{Partridge:1980vk}. It decays primarily
via $c\bar c$ annihilation into two gluons and is expected to have numerous hadronic decay
modes into two- or three-body hadronic charged and/or neutral final states, which,
unfortunately, seems not friendly for experimental studies at LHC experiments. Hence, though
struggling against the large background and the small efficiency at LHC for these problems,
the $B_c^+ \to \eta_c \pi^+$ decay is not yet observed presently. However, along with the
successful upgrade and resumed running of Large Hadron Collier-beauty (LHCb) detector since
2022, the ensuingly exciting results are forthcoming~\cite{Gao:2024lhc}. 
The continuously collected data from the upgraded LHCb detector will offer a precious
opportunity to advance $B_c$-meson physics into a precision era. 
It is therefore expected that a huge amount of data with superior quality can facilitate
a promising measurement of $B_c^+ \to \eta_c \pi^+$, for example, via $\eta_c \to p\bar p$
decay chain. Here, $p$ denotes a proton.

On the theoretical side, the $B_c^+ \to \eta_c \pi^+$ decay, as well as $B_c^+ \to \eta_c
\rho^+$, has been investigated in different approaches~\cite{Chang:1992pt,Anisimov:1998uk,Colangelo:1999zn,AbdElHady:1999xh,Verma:2001hb, Ebert:2003cn,Ivanov:2006ni,Hernandez:2006gt,Choi:2009ym,Naimuddin:2012dy,Qiao:2012hp,Rui:2014tpa,
Issadykov:2018myx,Nayak:2022qaq,Wu:2024gcq,Deng:2025znr}, while the branching ratios
(BRs) differ with a wide range of magnitude,
notably, ${\rm BR}(B_c^+ \to \eta_c \pi^+) \in [0.25, 4.22] \times
10^{-3}$ and ${\rm BR}(B_c^+ \to \eta_c \rho^+) \in [0.67, 13.16] \times 10^{-3}$, respectively.
These numerical results imply different understanding of QCD dynamics in the $B_c
\to \eta_c$ decays. The remarkable discrepancies suggest that our understanding of the
involved dynamics is far from complete and more investigations need to be carried out
in alternative approaches necessarily. Motivated by the general consistency between the
LHC measurements~\cite{LHCb:2024nlg,LHCb:2012ag,CMS:2014oqy,LHCb:2023fqn} and the
improved perturbative QCD (iPQCD) predictions~\cite{Liu:2023kxr,Liu:2025ipe} about the
ratios among the BRs of $B_c^+ \to J/\psi \pi^+$, $B_c^+ \to J/\psi \rho^+ (\to \pi^+\pi^0)$,
$B_c^+ \to J/\psi a_1(1260)(\to \pi^+\pi^-\pi^+)$, $B_c^+ \to \chi_{c1}(1P) \pi^+$, and
$B_c^+ \to \chi_{c2}(1P) \pi^+$, the present work will concentrate on the decays $B_c^+
\to \eta_c L^+$, where $L$ denotes the light mesons such as pseudoscalars ($P$)$- \pi$ and
$K$; vectors ($V$)$- \rho$ and $K^*$; axial-vectors ($A$)$- a_1(1260)$, $b_1(1235)$ and $K_1(1270,
1400)$; scalars ($S$)$- a_0(980, 1450)$ and $K_0^*(700, 1430)$ [$K_0^*(700)$ also known as
$\kappa$]; and tensors ($T$)$- a_2(1320)$ and $K_2^*(1430)$, respectively.

To the best of our knowledge, the nature of light scalars, especially those under or near
1~GeV, remains a long-standing puzzle in hadron physics. Recently, the CMS Collaboration
found strong evidence of $f_0(980)$ being a normal quark-antiquark state~\cite{CMS:2023rev},
even though the ALICE Collaboration supported the $K_0^*(700)$ being a four-quark state
\cite{ALICE:2023eyl}. Undoubtedly, more endeavors need to be devoted to this field.
The useful clues about the nature
of light scalars could be collected indirectly through probing their productions in the heavy
hadron decays, just like the $B \to f_0(980) K$ decays observed in the $B$-factory experiments
\cite{Abe:2002av,Aubert:2003mi}. So far, there
are two different scenarios to describe these scalar mesons $a_0(980, 1450)$ and $K_0^*(700,
1430)$ under the $q\bar q$ assignment~\cite{Cheng:2005nb}. In scenario 1 ($S1$), the scalar
mesons $a_0(980)$ and $K_0^*(700)$ are treated as the lowest-lying states, and those $a_0(1450)$
and $K_0^*(1430)$ are the first excited states correspondingly. And, in scenario 2 ($S2$),
the scalar mesons $a_0(1450)$ and $K_0^*(1430)$ are viewed as the ground states, while those
$a_0(980)$ and $K_0^*(700)$ might be the four-quark states. However, as
stressed in Ref.~\cite{Cheng:2005nb}, it is difficult in practice to make quantitative
predictions based on the four-quark or tetraquark picture for light scalars because the
calculations of decay constant and form factors of light scalars are beyond the conventional
quark model and the involved nonfactorizable contributions cannot be calculated in the
available QCD-based factorization framework. Moreover,
the productions of light scalars from
the vacuum in the $B_c^+ \to \eta_c S^+$ decays are expected to be highly suppressed
originating from the nearly zero vector decay constants $f_S$ [ actually, the vector
decay constants $f_S =0$ in the $SU(3)$ limit.]~\cite{Cheng:2005nb}.
It means that investigations on these decays must go beyond naive factorization.
Therefore, the predictions in this work are made on
the basis of two-quark model for light scalars within the iPQCD framework.

The $p$-wave light axial vectors have been investigated at both experimental and theoretical
aspects. However, our understanding about their nature is still far from complete~\cite{Du:2022nno}.
In the spectroscopy study~\cite{ParticleDataGroup:2024cfk},  $a_1(1260)$ and $b_1(1235)$ are
the $ 1^3\!P_1$ and $1^1\!P_1$ axial-vector states, respectively, carrying quantum numbers
$J^{PC} = 1^{++}$ and $J^{PC} = 1^{+-}$ correspondingly. Although it is very interesting that
the strange $K_1(1270)$ and
$K_1(1400)$ mesons are generally regarded as the mixtures of $1^3\!P_1$ state $K_{1A}$ and
$1^1\!P_1$ state $K_{1B}$ due to the $SU(3)$ flavor broken symmetry~\cite{Cheng:2011pb},
\beq
\left(
\begin{array}{c} |K_1(1270) \rangle \\ |K_1(1400) \rangle \\ \end{array} \right ) &=&
  \left( \begin{array}{cc}
\sin{\theta_{K}} & \hspace{0.28cm} \cos{\theta_{K}} \\
 \cos{\theta_{K}} & -\sin\theta_{K} \end{array} \right )
 \left( \begin{array}{c}  |K_{1A}\rangle\\ |K_{1B} \rangle \\ \end{array} \right )\;,
 \label{eq:mixture}
\eeq
with mixing angle $\theta_{K}$. The value of $\theta_K$ can be related to the masses of the
$K_1(1270)$ and $K_1(1400)$, to the strong decays of the $K_1(1270)$ and $K_1(1400)$, and
to rates of weak decays to final states involving the $K_1(1270)$ and $K_1(1400)$~\cite{Suzuki:1993yc,Blundell:1995au}.
Thus, the decays such as $B_c^+ \to
(c\bar c) K_1(1270, 1400)^+$ would be of great interest for exploring the information
of $\theta_K$. However, presently, there is no consensus on the value of the mixing angle $\theta_K$,
and the results from various approaches are still quite controversial; e.g., see a short
overview in Ref.~\cite{Liu:2024lph} ( and references therein). We therefore take both referenced
values, i.e., $\theta_{K_1} \approx 33^\circ$ and $58^\circ$~\cite{Cheng:2011pb,Shi:2023kiy}
into account in the related numerical calculations of this work.

By incorporating the finite charm quark mass effects into Sudakov resummation of the large
logarithmic corrections to wave functions through $k_T$ resummation at the next-to-leading-logarithm
accuracy~\cite{Liu:2020upy}, besides including them in the hard kernel, the iPQCD formalism
is now self-consistent for systematically studying the $B_c$-meson decays and $B$-meson decaying into
charmonia. So far, facilitated by the newly proposed transverse-momentum-dependent $B_c$-meson
wave function~\cite{Liu:2018kuo}, we have studied the $B_c^+ \to J/\psi M^+$~\cite{Liu:2023kxr}
and $B_c^+ \to \chi_{cJ} (P, V)^+$~\cite{Liu:2025ipe} decays, in which, the BRs predicted in
the iPQCD formalism are generally consistent with several available predictions in other
approaches. Furthermore, the resultant relative ratios such as ${\rm BR}(B_c^+ \to J/\psi
a_1(1260)^+ (\to \pi^+\pi^-\pi^+)) /{\rm BR}(B_c^+ \to J/\psi \pi^+)$, ${\rm BR}(B_c^+ \to
J/\psi \rho^+ (\to \pi^+\pi^0))/{\rm BR} (B_c^+ \to J/\psi \pi^+)$, ${\rm BR}(B_c^+ \to
\chi_{c2} \pi^+)/{\rm BR}(B_c^+ \to J/\psi \pi^+)$, etc. agree well with the current data
reported by the LHC experiments within theoretical uncertainties. In light of these successful
outputs, we shall analyze the decays $B_c^+ \to \eta_c L^+$ in the iPQCD formalism at leading
order in the strong coupling $\alpha_s$~\cite{Liu:2018kuo,Liu:2020upy}.
Inputting the BRs~\cite{ParticleDataGroup:2024cfk}
of strong decays $\eta_c \to p\bar p$ and $\eta_c \to
\pi^+\pi^- (\pi^+\pi^-, K^+ K^-, p\bar p)$, we additionally present the BRs of multibody
modes arising from $B_c^+ \to \eta_c (\pi, \rho)^+$ via resonance $\eta_c$ under the
narrow-width approximation. The (near-)future tests of our iPQCD predictions at
experimental facilities are expected to help us to further understand the QCD dynamics involved
in these $B_c$-meson decays, and even explore the inner structure of $\eta_c$ and related
light hadrons.

The rest of this paper is organized as follows. In Sec.~\ref{sec:form}, the formalism and the
perturbative calculations in association with factorization formulas of $B_c^+ \to \eta_c L^+$
are presented. The numerical results and phenomenological analyses are given in Sec.~\ref{sec:randd}.
Section~\ref{sec:summary} summarizes our main conclusions.


\section{ Formalism and perturbative calculations}
\label{sec:form}

For the $B_c^+ \to \eta_c L^+$ decays, the related weak effective Hamiltonian $H_{{\rm eff}}$
at the quark level can be written as~\cite{Buchalla:1995vs}
\beq
H_{\rm eff}\, &=&\, {G_F\over\sqrt{2}}
\biggl\{ V^*_{cb}V_{uq} [ C_1(\mu) O_1(\mu)
+C_2(\mu) O_2(\mu) ] \biggr\}+ {\rm H.c.}\;,
\label{eq:heff}
\eeq
with the Fermi constant $G_F=1.16639\times 10^{-5}{\rm GeV}^{-2}$, the Cabibbo-Kobayashi-Maskawa
(CKM) matrix elements $V$,
and the Wilson coefficients $C_i(\mu)$ at the renormalization scale $\mu$. The local four-quark
tree operators $O_1$ and $O_2$ are read as
\beq
O_1 \, &=&\,
\bar{q}_\alpha \gamma_\mu (1 - \gamma_5) u_\beta\; \bar{c}_\beta \gamma_\mu (1 - \gamma_5) b_\alpha \;,
\qquad
O_2 \, =\, \bar{q}_\alpha \gamma_\mu (1 - \gamma_5) u_\alpha\; \bar{c}_\beta \gamma_\mu (1 - \gamma_5) b_\beta \;,
\label{eq:operators}
\eeq
where $q$ denotes the light down quark $d (s)$ for the CKM-favored (-suppressed) $\Delta S =0\; (\Delta S =1)$ processes
with $S$ being strange number.

Similar to $B_c^+ \to J/\psi M^+$ decays~\cite{Liu:2023kxr}, the kinematics of $B_c^+ \to \eta_c L^+$
could be defined in the light-cone coordinates as
\beq
P_{1}&=& \frac{m_{B_c}}{\sqrt{2}}(1, 1, {\bf 0}_{T})\;,
\qquad
P_{2}= \frac{m_{B_c}}{\sqrt{2}}(1-r_{3}^{2}, r_{2}^{2}, {\bf 0}_{T})\;,
\qquad
P_{3}= \frac{m_{B_c}}{\sqrt{2}}(r_{3}^{2}, 1-r_{2}^{2}, {\bf 0}_{T})\;.
\label{eq:momenta}
\eeq
where the ratios $r_{2}=m_{L}/m_{B_c}$ and $r_{3}=m_{\eta_c}/m_{B_c}$ with $m_L (m_{\eta_c})$
being the light ($\eta_c$) meson mass, $P_1$ denotes the momentum carried
by $B_c$ meson in its rest frame, and $P_2$ and $P_3$ denotes the momenta carried by $L$ and $\eta_c$ mesons
moving along the plus and minus $z$ directions, respectively.
Thanks to conservation of the angular momentum, the possible polarization
vectors $\epsilon_{2L}$ ( here, the subscript $L$ stands for the longitudinal polarization, not to be
confused with the abbreviation $L$ of light mesons)
could be easily derived through the constraints $P_2 \cdot \epsilon_{2L} = 0$ and $\epsilon_{2L}^2 =-1$.
Notice that, if $L$ is a tensor meson, then a new longitudinal polarization vector could be constructed
similarly relative to those of vectors and axial vectors but with an additional factor $\sqrt{2/3}$~\cite{Berger:2000wt,Datta:2007yk,Liu:2017cwl,Liu:2023kxr}. Then, specifically, $\epsilon_{2L}=
\frac{1}{\sqrt{2(1-r_{3}^{2})} r_{2}}(1-r_{3}^{2}, -r_{2}^{2}, {\bf 0}_{T})$ for vectors and axial vectors,
while $\epsilon_{2L}= \frac{1}{\sqrt{3(1-r_{3}^{2})} r_{2}}(1-r_{3}^{2}, -r_{2}^{2}, {\bf 0}_{T})$ for tensors
in this work. The momenta of valence quarks in the initial- and final-state mesons are parametrized as
\beq
k_1 &=& (x_1 P_1^+, x_1 P_1^-, {\bf k}_{1T}) \;,
\qquad
k_2 = (x_2 P_2^+, x_2 P_2^-, {\bf k}_{2T}) \;,
\qquad
k_3 = (x_3 P_3^+, x_3 P_3^-, {\bf k}_{3T}) \;,
\eeq
where $x_i(i=1,2,3)$ is the corresponding momentum fraction.

The $B_c^+ \to \eta_c L^+$ decay amplitude in the iPQCD formalism can therefore be conceptually written as,
\beq
A(B_c^+ \to \eta_c L^+) &\sim &\int\!\! d x_1 d
x_2 d x_3 b_1 d b_1 b_2 d b_2 b_3 d b_3
\non && \cdot {\mathrm{Tr}}
\left [ C(t) \Phi_{B_c}(x_1, b_1) \Phi_{L}(x_2, b_2)
\Phi_{\eta_c}(x_3, b_3) H(x_i, b_i, t) e^{-S(t)} \right ]\;,
\label{eq:a2}
\eeq
where $b_i$ is the conjugate space coordinate of transverse momentum $k_{iT}$, $t$ is the largest running
energy scale in hard kernel $H(x_i,b_i,t)$, Tr denotes the trace over Dirac and $SU(3)$ color indices, $C(t)$
stands for the Wilson coefficients including the large logarithms $\ln (m_W/t)$~\cite{Keum:2000ph}, and
$\Phi$ is the wave function describing the hadronization of quark and antiquark to the meson. The Sudakov
factor $e^{-S(t)}$ arises from $k_T$ resummation, which provides a strong suppression on the long distance
contributions in the small $k_T$ (or large $b$) region~\cite{Botts:1989kf}. The detailed discussions for
$e^{-S(t)}$ can be easily found in the original Refs.~\cite{Botts:1989kf,Liu:2018kuo,Liu:2020upy}. Thus,
with Eq.~(\ref{eq:a2}), we can give the convoluted amplitudes of the decays $B_c^+ \to \eta_c L^+$ explicitly
through the evaluations of hard kernel $H(x_i,b_i,t)$ at leading order in the $\alpha_s$ expansion within
the iPQCD formalism.

The wave function for $B_c$ meson with a ``heavy-light" structure can generally be defined as~\cite{Liu:2018kuo,Keum:2000ph,Lu:2002ny}
\beq
\Phi_{B_c}(x, {\bf k}_T) &=& \frac{i }{\sqrt{2N_c}}
\biggl\{(\psl +m_{B_c})\gamma_5
 \phi_{B_c}(x, {\bf k}_T) \biggr\}\;,
\label{eq:def-bq}
\eeq
where
$P$ is the momentum of $B_c$ meson, $N_c =3$ is the color factor, and $x$ and ${\bf k}_T$ are the momentum
fraction and intrinsic transverse momentum of charm quark in the $B_c$ meson. Here, $\phi_{B_c}(x, {\bf k}_T)$
is the $B_c$-meson leading-twist distribution amplitude, whose explicit form in the impact ${\bf b}$ space
is~\cite{Liu:2018kuo}
\beq
\phi_{B_c}(x,{\bf b}) &=&   \frac{f_{B_c}}{2\sqrt{2 N_c } }
N_{B_c}x(1-x)\exp\left[-\frac{(1-x)m_c^2+xm_b^2}
{8\beta_{B_c}^2x(1-x)}\right]\exp\left[-2\beta_{B_c}^2x(1-x){\bf b}^2\right]\;,
\eeq
with the decay constant $f_{B_c} = 0.489 \pm 0.005$~GeV~\cite{Chiu:2007km},
the shape parameter $\beta_{B_c} = 1.0 \pm 0.1$~GeV~\cite{Liu:2018kuo}, and $m_c$ and $m_b$ being the charm and bottom quark masses.
Moreover, the normalization constant $N_{B_c}$ is fixed by the following relation:
\beq
\int_0^1 \phi_{B_c}(x, {\bf b}=0) dx \equiv\int_0^1 \phi_{B_c}(x) dx
&=& \frac{f_{B_c}}{2\sqrt{2 N_c} }\;.
\eeq

For the $\eta_c$ meson, its wave function has been studied within the nonrelativistic QCD
approach~\cite{Bondar:2004sv} and derived as
\beq
\Phi_{\eta_c}(x) &=& \frac{i}{\sqrt{2N_{c}}} \gamma_5
\biggl\{\psl\ \phi_{\eta_c}^{v}(x)
+m_{\eta_c} \phi_{\eta_c}^s(x)\biggl\}\;,
\label{eq:wf-etac}
\eeq
with $P$ and $m$ being the momentum and mass of $\eta_c$ and $x$ describing the charm-quark momentum
fraction in  $\phi^{v}_{\eta_c}(x)$ and $\phi^{s}_{\eta_c}(x)$ the twist-2 and twist-3 distribution
amplitudes,
\beq
\phi_{\eta_c}^{v}(x) &=& 9.58\frac{f_{\eta_c}}{2\sqrt{2N_{c}}}x(1-x){\cal C}(x) \;,
\qquad
\phi_{\eta_c}^{s}(x) = 1.97\frac{f_{\eta_c}}{2\sqrt{2N_{c}}} {\cal C} (x)
\;,
\label{eq:das-etac}
\eeq
where $f_{\eta_c} = 0.387 \pm 0.007$~\cite{Becirevic:2013bsa} is the decay constant
and the function ${\cal C}(x)$ reads
\beq
{\cal C}(x) &=& \biggl[(x(1-x))/(1-4x(1-x) (1-v^2))\biggl]^{1-v^2}\;,
\eeq
with $v^2 = 0.3$ standing for small relativistic corrections to the Coulomb wave functions.

The light-cone wave functions including distribution amplitudes for light pseudoscalars, scalars,
vectors, axial vectors, and tensors calculated in the QCD sum rules up to twist 3 have
been collected in Ref.~\cite{Liu:2023kxr} (and references therein). For simplicity, their
explicit expressions would no longer be presented in this work. The readers could refer to
Ref.~\cite{Liu:2023kxr} for detail.

\begin{figure}[!!htb]
\centering
\begin{tabular}{l}
\includegraphics[width=0.96\textwidth]{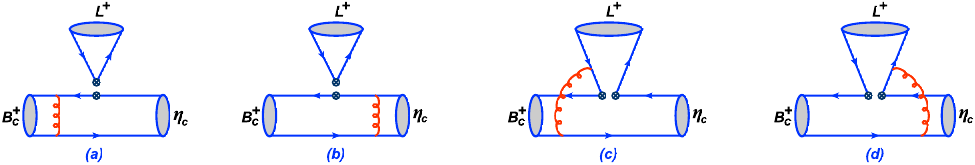}
\end{tabular}
\caption{ Leading-order Feynman diagrams for the decays $B_c^+ \to \eta_c L^+$
in the iPQCD formalism: (a) and (b) ((c) and (d)) contributing to the 
(non)factorizable-emission amplitudes. }
\label{fig:fig1}
\end{figure}

The leading-order Feynman diagrams for the decays $B_c^+ \to \eta_c L^+$ in the iPQCD formalism
are shown in Fig.~\ref{fig:fig1}. Similar to those in Ref.~\cite{Liu:2023kxr}, we use $F_{e}$ and $M_{e}$
to describe the factorizable emission and the nonfactorizable emission amplitudes induced by the
$(V-A)(V-A)$ operators. The $B_c^+ \to \eta_c L^+$ decay amplitude can thus be decomposed into
\beq
A(B_c^+ \to \eta_c L^+) &=& V_{cb}^* V_{uq} ( F_e \cdot f_{L} +  M_e )\;,
\label{eq:DecAmp}
\eeq
in which, the related factorization formulas are given explicitly as follows:
\begin{itemize}
    \item[(i)] {For $B_c^+ \to \eta_c (P, S)^+$ decays, }
\beq
F_e (P) &=& -8 \pi C_F m_{B_c}^4 \int_0^1 dx_1 dx_3
\int_0^\infty b_1db_1 b_3db_3 \phi_{B_c}(x_1,b_1) (r_3^2 -1)
\non && \times
\biggl\{  \left[ r_3 (r_b +2 x_3 -2) \phi_{\eta_c}^s(x_3)
-
(2 r_b + x_3 -1) \phi_{\eta_c}^v(x_3)  \right] h_a(x_1,x_3,b_1,b_3) E_f(t_a)
 \non &&
-
\left[2r_3(1+r_c -x_1)\phi_{\eta_c}^s(x_3) + (r^2_3 (x_1 -1)-r_c)\phi_{\eta_c}^v(x_3) \right] h_b(x_1,x_3,b_1,b_3)
E_f(t_b)  \biggr\} \;,
\label{eq:fe-P}
\eeq
where the ratios $r_b= m_b/m_{B_c}$ and $r_c = m_c/ m_{B_c}$. The hard function $h_i(x_i,
b_i)$ and the evolution function $E_f(t_i)$ could refer to those expressions in Ref.~\cite{Liu:2023kxr},

\beq
M_e (P) &=& - \frac{32}{\sqrt{6}}\pi C_F m_{B_c}^4 \int_0^1 dx_1 dx_2 dx_3
\int_0^\infty b_1db_1 b_2db_2 \phi_{B_c}(x_1,b_1) \phi_P^A(x_2) (r_3^2 -1)
\non & &
\times
\biggl\{\left[r_3 (x_3 -x_1)\phi_{\eta_c}^s(x_3) + (x_1+x_2-1)\phi_{\eta_c}^v(x_3) + r_3^2 (x_1 -x_2 -2 x_3 +1) \phi_{\eta_c}^v(x_3)
\right]
\non &&
\times
E_f(t_c) h_c(x_1,x_2,x_3,b_1,b_2)+\left[(x_2 +x_3 -2x_1+r_3^2(x_3-x_2))\phi_{\eta_c}^v(x_3)
\right.
\non &&
\left.
+ r_3 (x_1-x_3)\phi_{\eta_c}^s(x_3)\right]
h_d(x_1,x_2,x_3,b_1,b_2)E_f(t_d)\biggr\}\;,
\label{eq:nfe-P}
\eeq
 and
\beq
F_e (S) &=& - F_e (P) \;, \qquad   M_e (S) = - M_e (P)\;,
\label{eq:fe-nfe-S}
\eeq
but with the corresponding replacement of $\phi_P^A(x) \to \phi_S(x)$
in Eq.~(\ref{eq:fe-nfe-S}).

    \item[(ii)] {For $B_c^+ \to \eta_c (V, A, T)^+$ decays, }
\beq
F_e (V) & = & 8 \pi C_F m_{B_c}^4 \int_0^1 dx_1 dx_3
\int_0^\infty b_1db_1 b_3db_3 \phi_{B_c}(x_1,b_1) \sqrt{1 -r_3^2}
\non && \times
\biggl\{  \left[ r_3 (r_b +2 x_3 -2) \phi_{\eta_c}^s(x_3)
-
(2 r_b + x_3 -1) \phi_{\eta_c}^v(x_3)  \right] h_a(x_1,x_3,b_1,b_3) E_f(t_a)
 \non &&
 -
\left[2 r_3 (1+ r_c -x_1) \phi_{\eta_c}^s(x3) +(r^2_3 (x_1 -1)-r_c) \phi_{\eta_c}^v(x_3) \right] h_b(x_1,x_3,b_1,b_3)
E_f(t_b)  \biggr\} \;,
\label{eq:fe-V}
\eeq
\beq
M_e (V) &=& \frac{32}{\sqrt{6}}\pi C_F m_{B_c}^4 \int_0^1 dx_1 dx_2 dx_3
\int_0^\infty b_1db_1 b_2db_2 \phi_{B_c}(x_1,b_1) \phi_V(x_2) \sqrt{1-r_3^2}  \non & & \times
\biggl\{\left[(x_1+x_2-1)\phi_{\eta_c}^v(x_3) +r_3^2 (x_1 -x_2 - 2 x_3 +1) \phi_{\eta_c}^v(x3)
+r_3 (x_3 -x_1)\phi_{\eta_c}^s(x_3)\right]
\non &&
 E_f(t_c)  h_c(x_1,x_2,x_3,b_1,b_2)
+\left[(x_2+x_3-2 x_1+r_3^2(x_3-x_2))\phi_{\eta_c}^v(x_3)
\right.
\non &&
\left.
+ r_3 (x_1-x_3)\phi_{\eta_c}^s(x_3)\right] h_d(x_1,x_2,x_3,b_1,b_2)E_f(t_d)\biggr\}\;,
\label{eq:nfe-V}
\eeq
and
\beq
F_e (A) &=& - F_e (V) \;,  \qquad
F_e (T) = 0 \;,
\qquad
M_{e} (A) = - M_{e} (V) \;, \qquad
M_{e} (T) = \sqrt{\frac{2}{3}} M_{e} (V) \;.
\label{eq:Amp-at}
\eeq
associated with the replacements of $\phi_V(x) \to \phi_{A, T}(x)$
in (\ref{eq:Amp-at}) correspondingly.
\end{itemize}

The corresponding BR is then given, in the rest frame of a heavy $B_c$ meson, by
\beq
{\rm BR}(B_c^+ \to \eta_{c} L^+)&\equiv&
\tau_{B_c}\cdot \Gamma(B_c^+ \to \eta_c L^+)
= \tau_{B_c}\cdot\frac{G_{F}^{2}|\bf{P_c}|}{16 \pi m^{2}_{B_c} }
|A(B_c^+ \to \eta_c L^+)|^2\;,
\label{eq:BR-def}
\eeq
with the $B_c$-meson lifetime $\tau_{B_c}$ and the decay width $\Gamma$. Note that $|{\bf P}_c|\equiv
|{\bf P}_{L}| = |{\bf P}_{\eta_c}| = \sqrt{\lambda(m_{B_c}^2, m_{L}^2, m_{\eta_c}^2)}/(2 m_{B_c})$
is the momentum of either $L$ or $\eta_c$ meson in the final states,
with the ${\rm K\ddot{a}ll\acute{e}n}$ function $\lambda(x, y, z) = x^2 +y^2
+z^2 - 2 xy - 2 xz - 2 yz$~\cite{ParticleDataGroup:2024cfk}.

\section{Numerical Results and Discussions} \label{sec:randd}

In numerical calculations, central values of the input parameters will be used implicitly unless
otherwise stated. We adopt the relevant QCD scale~({\rm GeV}), masses~({\rm GeV}), and $B_c$-meson
lifetime ({\rm ps})~\cite{Keum:2000ph,ParticleDataGroup:2024cfk},
\beq
 \Lambda_{\overline{\mathrm{MS}}}^{(f=4)} &=& 0.250\; ,
 \qquad m_W = 80.41\;,
 \qquad m_{B_c} = 6.275 \;,
 \qquad m_{\eta_c}  = 2.98\;,
 \non
   \tau_{B_c} &=& 0.507\;,
 \qquad  m_b = 4.8 \;,
 \qquad  m_{c}= 1.5 \;.
\label{eq:mass}
\eeq
and the CKM matrix elements~\cite{ParticleDataGroup:2024cfk},
\beq
|V_{cb}| &=& 0.04182^{+0.00085}_{-0.00074}\;,
\quad
|V_{ud}| = 0.97435 \pm 0.00016\;,
\quad
|V_{us}| = 0.22500 \pm 0.00067\;.
\eeq

In the following context, we will classify the considered decays into two different groups
conventionally, that is, the decays $B_c^+ \to \eta_c (P, V)^+$
and the decays $B_c^+ \to \eta_c (A, S, T)^+$,
respectively, to present the related iPQCD
predictions and phenomenological insights.

\subsection{\boldmath  $B_c^+ \to \eta_c (P, V)^+$}
\label{ssec:fed}

The {\it CP}-averaged $B_c^+ \to \eta_c (\pi, K)^+$ BRs in the iPQCD formalism are presented as
\beq
{\rm BR}(B_c^+ \to \eta_c \pi^+)&=&
2.03
^{+0.51}_{-0.40}(\beta_{B_c})
^{+0.09}_{-0.08}(f_M)
^{+0.00}_{-0.00}(a_\pi)
^{+0.09}_{-0.07}(V_{cb})
\times 10^{-3} \;,
\label{eq:br-e-pi}
\\
{\rm BR}(B_c^+ \to \eta_c K^+) &=&
1.52
^{+0.39}_{-0.29}(\beta_{B_c})
^{+0.07}_{-0.06}(f_M)
^{+0.11}_{-0.10}(a_K)
^{+0.07}_{-0.05}(V_{cb})
\times 10^{-4}\;,
\eeq
where the uncertainties are dominated by the shape parameter $\beta_{B_c}$ from the $B_c$-meson
distribution amplitude.
Note that, though having a smaller decay constant $f_{\eta_c}\sim 0.387$ than $f_{J/\psi}\sim 0.405$
and the same leading-twist distribution amplitudes in both $\eta_c$ and $J/\psi$, the $B_c^+ \to \eta_c
\pi^+$ BR is still larger than the $B_c^+ \to J/\psi \pi^+$ one in the iPQCD formalism. In fact,
from the numerical results of decay amplitudes shown in Table~\ref{tab:DecAmps-ejp}, it is found
that ${\rm BR}(B_c^+ \to \eta_c \pi^+)$ and ${\rm BR}(B_c^+ \to J/\psi \pi^+)$ are
governed by highly different QCD dynamics, that
is, the former (latter) determined by the contributions arising from
the distribution amplitude of twist-3 $\phi_{\eta_c}^s(x)$ [twist-2 $\phi_{J/\psi}^L(x)$]
of $\eta_c (J/\psi)$, even with enhancement (reduction)
by that of twist-2 $\phi_{\eta_c}^v(x)$ [twist-3 $\phi_{J/\psi}^t(x)$].
Any precise measurements in various experiments to help understand the QCD
behavior of these $c\bar c$-mesons in depth are urgently demanded.

\begin{table}[htb]
\caption{
Decay amplitudes (in units of $10^{-3}$GeV$^{-3}$) of $B_c^+ \to \eta_c
\pi^+$ and $B_c^+ \to J/\psi \pi^+$ from different twists in the iPQCD formalism. For simplicity, only the central values are quoted for clarifications.}
\label{tab:DecAmps-ejp}
\begin{center}\vspace{-0.3cm}{\footnotesize
\begin{tabular}[t]{c||c|c|c|c}
\hline  \hline
Modes   & \multicolumn{2}{c|}{Decay Amplitudes ($F_e$)} & \multicolumn{2}{c}{Decay Amplitudes ($M_e$)} \\
 \hline \hline
 $B_c^+ \to \eta_c \pi^+$
& $\phi_{\eta_c}^v(x)$ &  $\phi_{\eta_c}^s(x)$
& $\phi_{\eta_c}^v(x)$ &  $\phi_{\eta_c}^s(x)$  \\
\hline
&$ 8.43 - {\it i} 8.56$
&$ -26.42 -{\it i} 113.08$
&$ 1.00 - {\it i} 2.61$
&$ 0.01 - {\it i} 0.01$\\
 \hline\hline
$B_c^+ \to J/\psi \pi^+$
& $\phi_{J/\psi}^L(x)$ &  $\phi_{J/\psi}^t(x)$
& $\phi_{J/\psi}^L(x)$ &  $\phi_{J/\psi}^t(x)$  \\
\hline
&$30.76 + {\it i} 101.07$
&$-12.83 -{\it i} 9.09$
&$-1.85 + {\it i} 4.21$
&$-0.24 -{\it i} 0.92$
\\
\hline \hline
\end{tabular}}
\end{center}
\end{table}

The ratio between ${\rm BR}(B_c^+ \to \eta_c \pi^+)$ and ${\rm BR}(B_c^+ \to J/\psi \pi^+)$ in the
iPQCD framework could help measure the mode $B_c^+ \to \eta_c \pi^+$ experimentally,
\beq
R_{\eta_c/{J/\psi}}^\pi &\equiv&
\frac{{\rm BR}(B_c^+ \to \eta_c \pi^+)}{{\rm BR}(B_c^+ \to J/\psi \pi^+)}
= 1.74^{+0.66}_{-0.50} \;,
\label{eq:r-e-j-pi}
\eeq
According to the LHCb data~\cite{LHCb:2014mvo}, that is, $\frac{\sigma(B_c^+)}{\sigma(B^+)}\times
{\rm BR}(B_c^+ \to J/\psi \pi^+) =(6.97\pm 0.13) \times 10^{-6}$~\cite{Liu:2025ipe} in the fiducial
region corresponding to the transverse momentum $0< p_T < 20$ GeV and the rapidity $2.0 < y < 4.5$,
and the assumption ${\rm BR}(B_c^+ \to J/\psi \pi^+) \sim {\cal O}(10^{-3})$, then the forthcoming
measurements of ${\rm BR}(B_c^+ \to \eta_c \pi^+)$ in the LHCb experiment might be
\beq
\frac{\sigma(B_c^+)}{\sigma(B^+)}\times {\rm BR}(B_c^+ \to \eta_c \pi^+) &\sim&
(1.41^{+0.37}_{-0.29}) \times 10^{-5}\;.
\eeq

Based on the data for $\eta_c$ strong decays to stable hadrons~\cite{ParticleDataGroup:2024cfk},
one can derive the multibody $B_c$-meson decays via $\eta_c$ resonance under the narrow-width approximation:
\begin{itemize}
\item[(1)]{$\eta_c \to p\bar p$ and $p\bar p \pi^+\pi^-$, }

The inputs of $\eta_c \to p\bar p$ and $p\bar p \pi^+\pi^-$ decays with
\beq
{\cal B}(\eta_c \to p\bar p) &=& (1.33 \pm 0.11)\times 10^{-3}\;, \qquad
{\cal B}(\eta_c \to p\bar p\pi^+\pi^-) = (3.7 \pm 0.5) \times 10^{-3}\;,
\label{eq:etac2proton}
\eeq
lead to the
three-body $B_c^+ \to \pi^+ \eta_c (\to p\bar p)$ and five-body $B_c^+ \to \pi^+\eta_c (\to p\bar p
\pi^+\pi^-)$ BRs as
\beq
{\rm BR}(B_c^+ \to \pi^+ \eta_c (\to p\bar p))&\equiv& {\rm BR}(B_c^+ \to \eta_c \pi^+)
\cdot {\cal B}(\eta_c \to p\bar p)
= (2.70^{+0.74}_{-0.59}) \times 10^{-6}\;,
\\
{\rm BR}(B_c^+ \to \pi^+ \eta_c (\to p\bar p \pi^+\pi^-))&\equiv& {\rm BR}(B_c^+ \to \eta_c \pi^+)
\cdot {\cal B}(\eta_c \to p\bar p\pi^+\pi^-)
= (7.51^{+2.21}_{-1.83}) \times 10^{-6}\;.
\eeq

\item[(2)]{$\eta_c \to 2(\pi^+\pi^-), 2(K^+K^-)$ and $\pi^+\pi^-K^+K^-$,}

The inputs of $\eta_c \to 2(\pi^+\pi^-)$, $\pi^+\pi^-K^+K^-$ and $2(K^+K^-)$ decays with
\beq
{\cal B}(\eta_c \to 2(\pi^+\pi^-)) &=& (9.6 \pm 1.5) \times 10^{-3}\;, \qquad
{\cal B}(\eta_c \to 2(K^+K^-)) = (1.4 \pm 0.4) \times 10^{-3}\;,\non
{\cal B}(\eta_c \to \pi^+\pi^-K^+K^-) &=& (8.3 \pm 1.8) \times 10^{-3}\;,
\label{eq:etac4pik}
\eeq
result in the
five-body $B_c^+ \to \pi^+\eta_c (\to 2(\pi^+\pi^-), (\pi^+\pi^-K^+K^-),
2(K^+K^-))$ BRs as
\beq
{\rm BR}(B_c^+ \to \pi^+ \eta_c (\to \pi^+\pi^-\pi^+\pi^-))&\equiv& {\rm BR}(B_c^+ \to \eta_c \pi^+)
\cdot {\cal B}(\eta_c \to \pi^+\pi^-\pi^+\pi^-)\non
&=& (1.95^{+0.59}_{-0.50}) \times 10^{-5}\;,\\
{\rm BR}(B_c^+ \to \pi^+ \eta_c (\to \pi^+\pi^-K^+K^-))&\equiv& {\rm BR}(B_c^+ \to \eta_c \pi^+)
\cdot {\cal B}(\eta_c \to \pi^+\pi^-K^+K^-)\non
&=& (1.68^{+0.57}_{-0.50}) \times 10^{-5}\;,\\
{\rm BR}(B_c^+ \to \pi^+ \eta_c (\to K^+K^-K^+K^-))&\equiv& {\rm BR}(B_c^+ \to \eta_c \pi^+)
\cdot {\cal B}(\eta_c \to K^+K^-K^+K^-)\non
&=& (2.84^{+1.10}_{-0.99}) \times 10^{-6}\;.
\eeq
\end{itemize}
In principle, the related measurements are previously difficult due to the large background and
the small efficiency at LHC.
However, these BRs around ${\cal O}(10^{-6})$ and above
with fully charged final states are expected to be probed in the near future, since the LHCb detector
has accomplished a successful upgrade in 2022.

The ratio between the BRs of $B_c^+ \to \eta_c K^+$ and $B_c^+ \to \eta_c \pi^+$ is given
theoretically in the iPQCD formalism as,
\beq
R^{\eta_c}_{K/\pi}&\equiv& \frac{{\rm BR}(B_c^+ \to \eta_c K^+)}
{{\rm BR}(B_c^+ \to \eta_c \pi^+)}=
(7.49
^{+0.54}_{-0.49})
\times 10^{-2}\;,
\label{eq:rkpi-th}
\eeq
which agrees well with the naive anticipation $(7.95 \pm 0.04) \times 10^{-2}$ within
uncertainties. That is, the decays $B_c^+ \to \eta_c (\pi, K)^+$ are indeed predominated
by the factorizable decay amplitudes. Meanwhile, the $a_1^K$-term induced $SU(3)$ flavor
symmetry-breaking effects
arising from leading-twist kaon distribution amplitude in the nonfactorizable
decay amplitudes lead to a slight deviation to the naive expectation, as stated in
Refs.~\cite{Liu:2023kxr,Liu:2025ipe}.

The $B_c^+ \to \eta_c V^+$ BRs in the iPQCD formalism can be read as
\beq
{\rm BR}(B_c^+ \to \eta_c \rho^+)&=&
5.37
^{+1.42}_{-1.08}(\beta_{B_c})
^{+0.24 }_{-0.24 }(f_M)
^{+0.00}_{-0.00}(a_\rho)
^{+0.22}_{-0.19}(V_{cb})
\times 10^{-3}  \;,
\\
{\rm BR}(B_c^+ \to \eta_c K^{*+})&=&
3.05
^{+0.81}_{-0.62}(\beta_{B_c})
^{+0.19 }_{-0.19 }(f_M)
^{+0.02}_{-0.02}(a_{K^*})
^{+0.12}_{-0.11}(V_{cb})
\times 10^{-4} \;,
\eeq
where the theoretical errors are also dominated mainly by the $B_c$-meson shape parameter $\beta_{B_c}$.
By employing Eqs.~(\ref{eq:etac2proton}) and~(\ref{eq:etac4pik}), the measurable BRs of
multibody $B_c^+\to \eta_c \rho^+$ decays through resonance state $\eta_c$ could be derived
under the narrow-width approximation as follows:
\beq
{\rm BR}(B_c^+ \to \eta_c (\to p \bar p) \rho^+) &\equiv& {\rm BR}(B_c^+ \to \eta_c \rho^+) \cdot
{\cal B}(\eta_c \to p\bar p) = (7.14^{+2.03}_{-1.60}) \times 10^{-6} \;, \\
{\rm BR}(B_c^+ \to \eta_c (\to p \bar p \pi^+ \pi^-) \rho^+) &\equiv& {\rm BR}(B_c^+ \to \eta_c \rho^+) \cdot
{\cal B}(\eta_c \to p\bar p \pi^+\pi^-) = (1.99^{+0.60}_{-0.49}) \times 10^{-5} \;,\\
{\rm BR}(B_c^+ \to \eta_c (\to \pi^+\pi^-\pi^+\pi^-) \rho^+) &\equiv& {\rm BR}(B_c^+ \to \eta_c \rho^+) \cdot
{\cal B}(\eta_c \to \pi^+\pi^-\pi^+\pi^-) \non
&=& (5.16^{+1.62}_{-1.34}) \times 10^{-5}\;,\\
{\rm BR}(B_c^+ \to \eta_c (\to \pi^+\pi^-K^+K^-) \rho^+) &\equiv& {\rm BR}(B_c^+ \to \eta_c \rho^+) \cdot
{\cal B}(\eta_c \to \pi^+\pi^-K^+K^-) \non
&=& (4.46^{+1.55}_{-1.34}) \times 10^{-5}\;,
\eeq
\beq
{\rm BR}(B_c^+ \to \eta_c (\to K^+K^-K^+K^-) \rho^+) &\equiv& {\rm BR}(B_c^+ \to \eta_c \rho^+) \cdot
{\cal B}(\eta_c \to K^+K^-K^+K^-) \non
&=& (7.52^{+2.97}_{-2.66}) \times 10^{-6}\;.
\eeq

The ratio between the $B_c^+ \to \eta_c \rho^+$ and $B_c^+ \to (J/\psi, \eta_c) \pi^+$ BRs could be
deduced as
\beq
\frac{{\rm BR}(B_c^+ \to \eta_c \rho^+)}{{\rm BR}(B_c^+ \to J/\psi \pi^+)}
&=&
4.59^{+1.77}_{-1.34}  \;,
\eeq
and
\beq
R_{\rho/\pi}^{\eta_c} &\equiv&
\frac{{\rm BR}(B_c^+ \to \eta_c \rho^+)}{{\rm BR}(B_c^+ \to \eta_c \pi^+)}=
2.65^{+0.02}_{-0.03}   \;,
\eeq
Here, the latter ratio $R_{\rho/\pi}^{\eta_c}$ is close to that obtained from the $B_c^+ \to
J/\psi (\pi, \rho)^+$ decays~\cite{Liu:2023kxr}.
Moreover, the ratio between the $B_c^+ \to \eta_c K^{*+}$ and
$B_c^+ \to \eta_c \rho^+$ BRs is
\beq
R^{\eta_c}_{K^*/\rho}&\equiv&
\frac{{\rm BR}(B_c^+ \to \eta_c K^{*+})}{{\rm BR}(B_c^+ \to \eta_c \rho^+)}
= (5.68
^{+0.11}_{-0.11})  \times 10^{-2} \;,
\label{eq:rrhkst-th}
\eeq
which matches well with the value $(5.75 \pm 0.03) \times 10^{-2}$ anticipated
by naive factorization within errors.

\begin{table}[hbt]
\caption{ Various predictions for ${\rm BR}(B_c^+ \to \eta_c \pi^+)$, ${\rm BR}(B_c^+ \to
\eta_c \pi^+)/{\rm BR}(B_c^+ \to J/\psi \pi^+)$ and ${\rm BR}(B_c^+ \to \eta_c K^+)/{\rm BR}
(B_c^+ \to \eta_c \pi^+)$ in this work and in the literature, respectively.}
\label{tab:Rpis-Refs}
\begin{center}\vspace{-0.3cm}{\tiny
\begin{tabular}[t]{c|c|c|c|c|c|c|c|c|c|c|c|c|c|c|c|c|c}
\hline\hline
Observables & This work &\cite{Chang:1992pt} & \cite{Anisimov:1998uk} &\cite{Colangelo:1999zn} & \cite{AbdElHady:1999xh} &\cite{Verma:2001hb}  &\cite{Ebert:2003cn} & \cite{Ivanov:2006ni} & \cite{Hernandez:2006gt}&
\cite{Choi:2009ym}\footnote{The results are calculated on the basis of Coulomb plus linear
confining and harmonic oscillator (in the parentheses) potentials.} & \cite{Naimuddin:2012dy}
& \cite{Qiao:2012hp}\footnote{These results are calculated with the non-relativistic QCD
approach at leading order in the strong coupling $\alpha_s$. And the corresponding next-to-leading-order
BR in $\alpha_s^2$ is $(5.19^{+0.70}_{-1.07}) \times 10^{-3}$.} & \cite{Rui:2014tpa} &\cite{Issadykov:2018myx} &\cite{Nayak:2022qaq} & \cite{Wu:2024gcq} & \cite{Deng:2025znr}\footnote{These results are
calculated with the QCD factorization approach in $\alpha_s$. And the corresponding next-to-next-to-leading-order BR in $\alpha_s^2$ is $(0.81^{+0.88}_{-0.56}) \times 10^{-3}$.}
\\
\hline
$10^{3}\cdot{\rm BR}(B_c^+ \to \eta_c \pi^+)$
& \Gre{$2.03^{+0.53}_{-0.41}$}& $1.80$ & $1.30$ & $0.26$ & $1.40$ & $9.30$ & $0.85$ & $1.90$ & $0.94$ & $0.91 (1.16)$
& $0.34$  & $ 2.95$ & $2.98^{+1.24}_{-1.05}$  &$1.89^{+0.37}_{-0.37}$ & $0.397$
& $1.40^{+0.40}_{-0.40}$ & $0.79^{+0.86}_{-0.55}$
\\
\hline
$\frac{{\rm BR}(B_c^+ \to \eta_c \pi^+)}{{\rm BR}(B_c^+ \to J/\psi \pi^+)}
$
& \Gre{$1.74^{+0.66}_{-0.50}$}& $1.00$ & $1.78$& $0.19$ & $1.27$  & $2.07$& $1.39$& $1.12$ & $1.24$  & $\dots$
& $1.00$ & $ 1.33$
& $1.28^{+0.69}_{-0.56}$
& $1.87^{+0.52}_{-0.52}$ & $1.02$
& $0.56^{+0.21}_{-0.21}$
& $0.95^{+1.04}_{-0.68}$
 \\
\hline
$10^2\cdot R^{\eta_c}_{K/\pi}
$
&\Gre{$7.49^{+0.54}_{-0.49}$} & $7.78$ & $10.00$ & $8.00$  & $7.86$  & $5.05$ & $8.24$ & $7.89$ & $7.98$ & $8.13
(8.10)$ & $8.82$ & $ 7.12$ & $8.05^{+0.68}_{-0.99}$ & $7.94^{+0.02}_{-0.05}$
& $7.81$ & $7.86^{+0.14}_{-0.08}$ & $ 7.51^{+0.02}_{-0.00}$
\\
\hline \hline
\end{tabular}}
\end{center}
\end{table}

\begin{table}[hbt]
\caption{ Same as Table~\ref{tab:Rpis-Refs} but for $B_c^+ \to \eta_c (\rho, K^*)^+$.}
\label{tab:Rrhos-Refs}
\begin{center}\vspace{-0.3cm}{\tiny
\begin{tabular}[t]{c|c|c|c|c|c|c|c|c|c|c|c|c|c|c|c|c|c}
\hline\hline
Observables&This work &\cite{Chang:1992pt}& \cite{Anisimov:1998uk}&\cite{Colangelo:1999zn}& \cite{AbdElHady:1999xh} &\cite{Verma:2001hb}  &\cite{Ebert:2003cn} & \cite{Ivanov:2006ni} & \cite{Hernandez:2006gt}
& \cite{Choi:2009ym} & \cite{Naimuddin:2012dy} &\cite{Qiao:2012hp}
& \cite{Rui:2014tpa} &\cite{Issadykov:2018myx} &\cite{Nayak:2022qaq} & \cite{Wu:2024gcq}
& \cite{Deng:2025znr}
\\
\hline
$10^{3}\cdot{\rm BR}(B_c^+ \to \eta_c \rho^+)$
&\Gre{$5.37^{+1.46}_{-1.12}$} &$4.90$& $3.00$ & $0.67$ & $3.30$ &$370.0$ & $2.10$ & $4.50$ & $2.40$  & $2.57 (3.24)$
& $1.06$ &$ 7.89$ & $9.83^{+3.33}_{-2.59}$ & $5.18^{+1.04}_{-1.04}$  & $1.24$
& $3.80^{+0.10}_{-0.10}$ & $2.15^{+2.23}_{-1.46}$
\\
\hline
$\frac{{\rm BR}(B_c^+ \to \eta_c \rho^+)}{{\rm BR}(B_c^+ \to J/\psi \pi^+)}
$
& \Gre{$4.59^{+1.77}_{-1.34}$} &$2.72$ & $4.11$ & $0.52$ & $3.0$ &$82.2$ & $3.44$ & $2.65$ & $3.16$  &$\dots$ & $3.12$
& $3.55$ & $4.22^{+2.05}_{-1.57}$ & $5.13^{+1.45}_{-1.45}$ & $3.18$
& $1.52^{+0.37}_{-0.37}$ & $2.60^{+2.71}_{-1.80}$
 \\
\hline
$10^2\cdot R^{\eta_c}_{K^*/\rho}
$
& \Gre{$5.68^{+0.11}_{-0.11}$} & $5.10$ & $7.00$ &$5.97$  & $5.45$ & $3.70$ & $5.24$& $5.56$ & $5.42$ & $5.06 (5.25)$
& $5.66$ & $ 5.20$ &$5.80^{+0.52}_{-0.45}$ & $5.60^{+0.03}_{-0.04}$ &$5.24$
& $5.53^{+1.14}_{-1.48}$ & $5.12^{+0.15}_{-0.06}$
\\
\hline \hline
\end{tabular}}
\end{center}
\end{table}

Finally, as aforementioned in the Introduction, the decays $B_c^+ \to \eta_c (P, V)^+$ have been
studied extensively with various $B_c \to \eta_c$ form factors in different approaches~\cite{Chang:1992pt,Anisimov:1998uk,Colangelo:1999zn,AbdElHady:1999xh,Verma:2001hb, Ebert:2003cn,Ivanov:2006ni,Hernandez:2006gt,Choi:2009ym,Naimuddin:2012dy,Qiao:2012hp,Rui:2014tpa,
Issadykov:2018myx,Nayak:2022qaq,Wu:2024gcq,Deng:2025znr}. However,
the obtained values of BRs and relative ratios have a wide spread, as explicitly
presented in Tables~\ref{tab:Rpis-Refs} and~\ref{tab:Rrhos-Refs}.
Based on the numerical results in the literature and this work, some remarks are in order:
(a) For the absolute BRs of $B_c^+ \to \eta_c (\pi, \rho)^+$, as seen in the first line of Tables~\ref{tab:Rpis-Refs} and~\ref{tab:Rrhos-Refs}, it is clear that, within $1\sigma$ errors,
our iPQCD predictions are consistent with those in Refs.~\cite{Chang:1992pt,Ivanov:2006ni,Rui:2014tpa,Issadykov:2018myx,Wu:2024gcq,Deng:2025znr}
for ${\rm BR}(B_c^+ \to \eta_c \pi^+)$ and with those
in Refs.~\cite{Chang:1992pt,Ivanov:2006ni,Issadykov:2018myx,Deng:2025znr} for ${\rm BR}(B_c^+ \to
\eta_c \rho^+)$, respectively. We can also easily find that the results of BRs
for both $B_c^+ \to \eta_c \pi^+$ and $B_c^+ \to \eta_c \rho^+$ decays given in
Refs.~\cite{Colangelo:1999zn,Verma:2001hb,Ebert:2003cn,Hernandez:2006gt,Naimuddin:2012dy,Nayak:2022qaq}
are inconsistent with ours, even within $2\sigma$ errors.
Needless to say, there are significant discrepancies in the understanding of the transition $B_c \to \eta_c$ across different frameworks.
(b) For the relative ratios ${\rm BR}(B_c^+ \to \eta_c (\pi, \rho)^+)/{\rm BR}(B_c^+ \to J/\psi \pi^+)$, as collected in the second line of Tables~\ref{tab:Rpis-Refs}
and~\ref{tab:Rrhos-Refs}, it is found that, within $1\sigma$ errors, our iPQCD predictions
are consistent with those in Refs.~\cite{Anisimov:1998uk,AbdElHady:1999xh,Verma:2001hb,Ebert:2003cn,Hernandez:2006gt,Qiao:2012hp,Rui:2014tpa,
Issadykov:2018myx,Deng:2025znr} for ${\rm BR}(B_c^+ \to \eta_c \pi^+)/{\rm BR}(B_c^+ \to J/\psi \pi^+)$
and with those in Refs.~\cite{Anisimov:1998uk,Ebert:2003cn,Qiao:2012hp,Rui:2014tpa,
Issadykov:2018myx,Deng:2025znr} for ${\rm BR}(B_c^+ \to \eta_c \rho^+)/{\rm BR}(B_c^+ \to J/\psi \pi^+)$, respectively.
(c) As shown in the third line of Tables~II and~III, within $1\sigma$ errors,
our predictions for $R^{\eta_c}_{K/\pi}$ are consistent with most literature results
except those in Refs.~\cite{Anisimov:1998uk,Verma:2001hb, Ebert:2003cn,Naimuddin:2012dy}, 
while for $R^{\eta_c}_{K^*/\rho}$ they agree only
with those in Refs.~\cite{AbdElHady:1999xh,Ivanov:2006ni,Naimuddin:2012dy,Rui:2014tpa,
Issadykov:2018myx,Wu:2024gcq}.

It is worth noting that all our numerical results, such as individual BRs and various relative
ratios, for the decays $B_c^+ \to \eta_c (P, V)^+$ in the iPQCD formalism agree well with
those calculated in covariant confined quark model~\cite{Issadykov:2018myx}
within $1\sigma$ uncertainties. To evaluate the reliability of the available theoretical
calculations across different models, future precise measurements are highly essential.

\subsection{\boldmath  $B_c^+ \to \eta_c (A, S, T)^+$}
\label{ssec:nfed}

In this subsection, we present the predictions of the decays
$B_c^+ \to \eta_c (A, S, T)^+ $ for the first time within the QCD-based factorization framework,
which could provide references for understanding the structure of these considered
light hadrons in the future.

First of all, for the $B_c^+ \to \eta_c a_1(1260)^+$ channel,
its BR in the iPQCD formalism is read as
\beq
{\rm BR}(B_c^+ \to \eta_c a_1(1260)^+)&=&
6.91
^{+1.84}_{-1.41}(\beta_{B_c})
^{+0.66}_{-0.64}(f_M)
^{+0.00}_{-0.00}(a_{a_1})
^{+0.29}_{-0.24}(V_{cb})
\times 10^{-3} \;,
\eeq
associated with the relevant ratios,
\beq
\frac{{\rm BR}(B_c^+ \to \eta_c a_1(1260)^+)}{{\rm BR}(B_c^+ \to J/\psi \pi^+)}
&=&
5.91^{+2.33}_{-1.81}  \;,
\eeq
and
\beq
R_{a_1/\rho}^{\eta_c} &\equiv&
\frac{{\rm BR}(B_c^+ \to \eta_c a_1(1260)^+)}{{\rm BR}(B_c^+ \to \eta_c \rho^+)}
= 1.29^{+0.51}_{-0.40}
\;,
  \qquad
R_{a_1/\pi}^{\eta_c} \equiv
\frac{{\rm BR}(B_c^+ \to \eta_c a_1(1260)^+)}{{\rm BR}(B_c^+ \to \eta_c \pi^+)}
= 3.40^{+1.32}_{-1.03}  \;.
\eeq
In the above results, the uncertainties are mainly dominated by the variations of
shape parameter $\beta_{B_c}$ in the $B_c$-meson
distribution amplitude. 
Future precise measurements will provide valuable information to aid in the study
of QCD dynamics among the states pion, $\rho$, and $a_1(1260)$,
which involve the same quark components.

For the decay $B_c^+ \to \eta_c b_1(1235)^+$, in principle, different from the $B_c^+ \to
\eta_c (P, V)^+$ modes, its factorizable-emission contributions are generally suppressed due to
the tiny zeroth Gegenbauer moment $a_{0,b_1}^\parallel$ in the longitudinal leading-twist
$b_1(1235)^+$ distribution amplitude~\cite{Yang:2005gk}, while its iPQCD BR is read as
\beq
{\rm BR}(B_c^+ \to \eta_c b_1(1235)^+)&=&
7.88
^{+2.52}_{-1.86}(\beta_{B_c})
^{+0.80}_{-0.75}(f_M)
^{+3.07}_{-2.57}(a_{b_1})
^{+0.32}_{-0.28}(V_{cb})
\times 10^{-4} \;.
\eeq
accompanied by the two ratios
\beq
\frac{{\rm BR}(B_c^+ \to \eta_c b_1(1235)^+)}{{\rm BR}(B_c^+ \to J/\psi \pi^+)}
&=&
0.67^{+0.39}_{-0.31}      \;,
     \qquad
R_{b_1/\pi}^{\eta_c} \equiv
\frac{{\rm BR}(B_c^+ \to \eta_c b_1(1235)^+)}{{\rm BR}(B_c^+ \to \eta_c \pi^+)}
= 0.39^{+0.22}_{-0.18}      \;,
\eeq
in which theoretical uncertainties of these numerical results are dominantly induced
by the variations of shape parameter $\beta_{B_c}$ and Gegenbauer moment $a_{b_1}$ in the
leading-twist distribution amplitudes of $B_c$ and $b_1(1235)$, respectively.
Notice that, the BRs of $B_c^+ \to \eta_c b_1(1235)^+$ and $B_c^+ \to J/\psi b_1(1235)^+$ in the iPQCD formalism indicate
an interesting relation, namely, ${\rm BR}(B_c^+ \to \eta_c b_1(1235)^+) \simeq
{\rm BR}(B_c^+ \to J/\psi b_1(1235)^+) \sim {\cal O}(10^{-3})$ within uncertainties,
even though the latter BR includes three kinds of polarization contributions~\cite{Liu:2023kxr}.
These values could be
tested in the near-future experiments to help understand the QCD dynamics in
the related decays, as well as in $b_1(1235)$.

The $B_c^+ \to \eta_c K_1(1270, 1400)^+$ BRs predicted in the iPQCD formalism
with different $\theta_K$ are as follows,
\beq
{\rm BR}(B_c^+ \to \eta_c K_1(1270)^+)&=&
\left\{ \begin{array}{ll}
2.57
^{+0.63}_{-0.50}(\beta_{B_c})
^{+0.23}_{-0.22}(f_M)
^{+0.71}_{-0.63}(B_{K_1})
^{+0.11}_{-0.09}(V_{cb})
\times 10^{-4}
& \vspace{0.12cm}
\\
4.00
^{+1.01}_{-0.78}(\beta_{B_c})
^{+0.40}_{-0.38}(f_M)
^{+0.58}_{-0.54}(B_{K_1})
^{+0.16}_{-0.14}(V_{cb})
\times 10^{-4}
& \end{array} \right.,
\label{eq:br-k12}
\\
{\rm BR}(B_c^+ \to \eta_c K_1(1400)^+)&=&
\left\{ \begin{array}{ll}
2.10
^{+0.63}_{-0.45}(\beta_{B_c})
^{+0.27}_{-0.26}(f_M)
^{+0.37}_{-0.35}(B_{K_1})
^{+0.08}_{-0.08}(V_{cb})
\times 10^{-4}
& \vspace{0.12cm}
\\
6.70
^{+2.48}_{-1.68}(\beta_{B_c})
^{+0.78}_{-0.73}(f_M)
^{+2.92}_{-2.12}(B_{K_1})
^{+0.27}_{-0.24}(V_{cb})
\times 10^{-5}
& \end{array} \right.,
\label{eq:br-k14}
\eeq
where the first (second) entry in Eqs.~(\ref{eq:br-k12}) and~(\ref{eq:br-k14}) corresponds to
$\theta_{K} = 33^\circ\ (58^\circ)$.
The similar patterns to the BRs also appear in the following observables for related modes trivially:
\beq
\frac{{\rm BR}(B_c^+ \to \eta_c K_1(1270)^+)}{{\rm BR}(B_c^+ \to J/\psi \pi^+)}
&=&
\left\{ \begin{array}{ll}
0.22^{+0.10}_{-0.08}
& \vspace{0.12cm} \\
0.34^{+0.14}_{-0.11}
&  \\ \end{array} \right.,
\qquad
\frac{{\rm BR}(B_c^+ \to \eta_c K_1(1400)^{+})}{{\rm BR}(B_c^+ \to J/\psi \pi^+)}
=
\left\{ \begin{array}{ll}
0.18^{+0.08}_{-0.07}
& \vspace{0.12cm} \\
0.06^{+0.04}_{-0.03}
&  \\ \end{array} \right..
\eeq
In the above numerical results, the dominant errors arise from the uncertainties of shape
parameter $\beta_{B_c}$ and Gegenbauer moments in the distribution amplitudes of $B_c$, and
$K_{1A}$ and $K_{1B}$ states, respectively. The
$B_c^+ \to \eta_c K_1(1270, 1400)^+$ BRs indicate that ${\rm BR}(B_c^+ \to
\eta_c K_1(1270)^+)$ is very consistent with ${\rm BR}(B_c^+ \to \eta_c K_1(1400)^+)$ at
$\theta_{K} \sim 33^\circ$ within uncertainties, while ${\rm BR}(B_c^+ \to \eta_c K_1(1270)^+)$
is significantly larger than ${\rm BR}(B_c^+ \to \eta_c K_1(1400)^+)$ at $\theta_{K} \sim
58^\circ$ with a factor around 6. It means that a destructive interference between $B_c^+
\to \eta_c K_{1A}^+$ and $B_c^+ \to \eta_c K_{1B}^+$ occurs significantly in $B_c^+
\to \eta_c K_1(1400)^+$ at $\theta_{K} \sim 58^\circ$. In other words, future measurements
testing these $B_c^+ \to \eta_c K_1(1270, 1400)^+$ BRs with clearly different results would
lead to a better determination of $\theta_{K}$. A verified $\theta_K$ value is key to further
guide the theoretical predictions with good precision.

It is worth emphasizing that the determination of $\theta_{K}$ with definite value is
highly important because, if $\theta_{K}$ could be determined unambiguously, it could further
help constrain the mixing between $f_{1}(1285)(h_1(1170))$ and $f_1(1420)(h_1(1450))$
promisingly~\cite{Cheng:2011pb}. The ratios between the $B_c^+ \to \eta_c K_1(1270)^+$ and
$B_c^+ \to \eta_c K_1(1400)^+$ BRs are then derived to provide necessary reference for
constraining the magnitude of $\theta_{K}$,
\beq
\frac{{\rm BR}(B_c^+ \to \eta_c K_1(1400)^+)}{{\rm BR}(B_c^+ \to \eta_c K_1(1270)^+)}
&=&
\left\{ \begin{array}{ll}
0.82^{+0.09}_{-0.08}
& \vspace{0.12cm} \\
0.17^{+0.04}_{-0.04}
&  \\ \end{array} \right..
\eeq

The above BRs and the associated ratios predicted in the iPQCD formalism would be helpful to
explore the QCD dynamics in the considered axial vectors, especially in $K_1(1270, 1400)$.

Next, we turn to the $B_c^+ \to \eta_c S^+$ decays.
As stressed in the Introduction, relative to the above $B_c^+ \to \eta_c (P, V)^+$ decays,
the nearly zero vector decay constant $f_S$ for light scalars
will lead to the significantly suppressed factorizable-emission contributions in the $B_c^+ \to \eta_c S^+$
channels.
According to the categorization of light scalar mesons in two scenarios, the {\it CP}-averaged
$B_c^+ \to \eta_c S^+$ BRs in the iPQCD formalism are given as
\beq
{\rm BR}(B_c^+ \to \eta_c a_0(980)^+)&=&
1.02
^{+0.44}_{-0.30}(\beta_{B_c})
^{+0.24}_{-0.21}(f_M)
^{+0.04}_{-0.05}(B_i)
^{+0.04}_{-0.04}(V_{cb})
\times 10^{-7}\;,
\\
{\rm BR}(B_c^+ \to \eta_c \kappa^+)&=&
7.20
^{+3.16}_{-2.07}(\beta_{B_c})
^{+1.82}_{-1.52}(f_M)
^{+0.49}_{-0.46}(B_i)
^{+0.30}_{-0.25}(V_{cb})
\times 10^{-6}\;,
\eeq
and
\beq
{\rm BR}(B_c^+ \to \eta_c a_0(1450)^+) &=&
\left\{ \begin{array}{ll}
6.79
^{+1.61}_{-1.39}(\beta_{B_c})
^{+4.53}_{-1.40}(f_M)
^{+1.13}_{-0.65}(B_i)
^{+0.27}_{-0.24}(V_{cb})
\times 10^{-9}
&
\vspace{0.12cm}
\\
2.15
^{+0.83}_{-0.57}(\beta_{B_c})
^{+1.00}_{-1.00}(f_M)
^{+0.05}_{-0.04}(B_i)
^{+0.09}_{-0.07}(V_{cb})
\times 10^{-7}
&  \\ \end{array} \right.,
\label{eq:br-a014}
\\
{\rm BR}(B_c^+ \to \eta_c {K_0^{*}}(1430)^{+}) &=&
\left\{ \begin{array}{ll}
6.68
^{+1.70}_{-1.38}(\beta_{B_c})
^{+3.39}_{-2.06}(f_M)
^{+0.21}_{-0.16}(B_i)
^{+0.28}_{-0.23}(V_{cb})
\times 10^{-7}
& \vspace{0.12cm}
\\
1.29
^{+0.50}_{-0.34}(\beta_{B_c})
^{+0.75}_{-0.47}(f_M)
^{+0.04}_{-0.04}(B_i)
^{+0.06}_{-0.04}(V_{cb})
\times 10^{-5}
&  \\ \end{array} \right.,
\label{eq:br-k014}
\eeq
where the first (second) entry in Eqs.~(\ref{eq:br-a014}) and~(\ref{eq:br-k014}) corresponds to
$S1 (S2)$ and the dominant
errors come mainly from the $B_c$-meson shape parameter $\beta_{B_c}$ and from the scalar
decay constant $\bar f_{S}$ and the Gegenbauer moments $B_{i} (i=1,3)$ of scalar mesons,
respectively. The future precise constraints from experimental measurements and Lattice
QCD calculations on these hadronic parameters are urgently demanded for theoretical
predictions with good precision.
Nevertheless, since the LHC experiments can measure the $B_c$ decays with a branching ratio
at $10^{-6}$ level~\cite{Descotes-Genon:2009eui}, then the BRs $(7.20^{+3.69}_{-2.62})
\times 10^{-6}$ for $B_c^+ \to \eta_c \kappa^+$ and $(6.68^{+3.81}_{-2.50})
\times 10^{-7} [(1.29^{+0.91}_{-0.58})
\times 10^{-5}]$ for $B_c^+ \to \eta_c {K_0^*}(1430)^+$ are expected to be probed in the
newly upgraded LHCb experiment. Furthermore, the experimental confirmations on the evident
order difference of magnitudes in the predicted BRs of $B_c^+ \to \eta_c {K_0^*}(1430)^+$
could potentially provide important clues to ascertain the preferred scenario for
understanding the scalar $K_0^*(1430)$.

In principle, the above-mentioned large uncertainties from nonperturbative inputs tend to be
greatly cancelled by relative ratios between the relevant BRs. Then, the ratios between the
$\Delta S=1$ and $\Delta S=0$ BRs in $B_c^+ \to \eta_c S^+$ are easily given as,
\beq
\frac{{\rm BR}(B_c^+ \to \eta_c \kappa^+)}{{\rm BR}(B_c^+ \to  \eta_c a_0(980)^+)}
&=&
(0.71^{+0.02}_{-0.01}) \times 10^2
\;,
\qquad
\frac{{\rm BR}(B_c^+ \to \eta_c {K_0^*}(1430)^+)}{{\rm BR}(B_c^+ \to \eta_c a_0(1450)^+)}
=
\left\{ \begin{array}{ll}
(0.98^{+0.08}_{-0.16}) \times 10^{2}
& \vspace{0.12cm} \\
(0.60^{+0.11}_{-0.01}) \times 10^{2}
&  \\ \end{array} \right..
\eeq
It seems that the errors in these results induced by the hadronic parameters are indeed
canceled to a great extent. However, it is surprisingly noted that, drastically different
from the values presented in Eqs.~(\ref{eq:rkpi-th}) and~(\ref{eq:rrhkst-th}), the above
two ratios are significantly large near ${\cal O}(10^{2})$ within uncertainties, even though already
with a known factor of $|V_{us}/V_{ud}|^2 \sim 0.05$. Moreover, they are also
remarkably larger than those in the $B_c^+ \to J/\psi S^+$ decays correspondingly within the iPQCD
framework, for detail, see Eq.~(61) in Ref.~\cite{Liu:2023kxr}.

\begin{table}[htb]
\caption{
Decay amplitudes (in units of $10^{-3}$GeV$^{-3}$) of $B_c^+ \to
\eta_c\; (J/\psi)\; S^+$ in the iPQCD formalism.
The upper (lower) entry corresponds to the scalars
$a_0(1450)^+$ and ${K_0^*}(1430)^+$ in scenario 1 (2) at every line. For simplicity,
only the central values are quoted for clarifications.}
\label{tab:DecAmps}
\begin{center}\vspace{-0.3cm}{\footnotesize
\begin{tabular}[t]{c||c|c}
\hline  \hline
Modes   & Decay Amplitudes ($F_e$) & Decay Amplitudes ($M_e$) \\
\hline \hline
 $B_c^+ \to \eta_c\; (J/\psi)\; a_0(980)^+$
&$ - 0.32 - {\it i} 1.88\; (-0.11 - {\it i} 0.73)$
&$  0.59 +{\it i} 2.76\; (-48.13 + {\it i} 52.70)$
\\
\hline
$B_c^+ \to \eta_c\; (J/\psi)\; \kappa^+$
&$ 0.46 + {\it i} 10.09\; (-1.17 - {\it i} 7.60)$
&$1.51 -{\it i} 2.63\; (-7.87 + {\it i} 12.70)$
\\
\hline \hline
$B_c^+ \to \eta_c\; (J/\psi)\; a_0(1450)^+$
&$\begin{array}{cc}
 0.02 + {\it i} 0.13\; (0.09 + {\it i} 0.63)
\\
 0.09 + {\it i} 1.41\; (-0.15 - {\it i} 0.97)
\end{array}$
&$\begin{array}{cc}
- 0.14 + {\it i} 0.08\; (-68.32 + {\it i} 33.43)
\\
0.17 -{\it i} 0.05\; (-44.16 + {\it i} 7.08)
\end{array}$
\\
\hline
$B_c^+ \to \eta_c\; (J/\psi)\; {K_0^*}(1430)^+$
&$\begin{array}{cc}
 0.21 + {\it i} 1.60\; (1.07 + {\it i} 7.06)
\\
 0.71 + {\it i} 11.01\; (-1.60 - {\it i} 10.60)
\end{array}$
&$\begin{array}{cc}
- 1.04 + {\it i} 0.68\; (-11.39 + {\it i} 6.47)
\\
1.41 -{\it i} 0.53\; (-10.30 + {\it i} 2.90)
\end{array}$
\\
\hline \hline
\end{tabular}}
\end{center}
\end{table}

To understand this peculiar feature in the $B_c^+ \to \eta_c S^+$ decays, the related
decay amplitudes calculated in the iPQCD formalism are presented explicitly in Table~\ref{tab:DecAmps}.
For effective and clear clarifications, we also quote the available $B_c^+ \to J/\psi S^+$ decay amplitudes~\cite{Liu:2023kxr} and present them in Table~\ref{tab:DecAmps} simultaneously.
Evidently, the originally constructive interferences to the $B_c^+ \to J/\psi S^+$ BRs between
the nonfactorizable-emission diagrams in Figs.~\ref{fig:fig1}(c) and
\ref{fig:fig1}(d) with antisymmetric leading-twist distribution amplitude of light scalars do
not appear in the $B_c^+ \to \eta_c S^+$ modes. In contrast, the sharply destructive interferences
result in much smaller decay amplitudes as exhibited in the third column of
Table~\ref{tab:DecAmps}. Furthermore, the destructions in
$S1$ are heavier than those in $S2$ for the $B_c^+ \to \eta_c (a_0(1450), K_0^*(1430))^+$ decays.
The situation is very contrary to that in the $B_c^+ \to J/\psi S^+$ decays~\cite{Liu:2023kxr}.
Due to allowed $SU(3)$ symmetry-breaking effects,
the comparably large factorizable-emission contributions, although evidently suppressed relative
to those of $B_c^+ \to \eta_c (P, V)^+$, lead to further significant destructions between
the decay amplitudes of factorizable-emission and nonfactorizable-emission topologies in these
$B_c^+ \to \eta_c S^+$ decays. However, it should be stressed that, due to the light quark masses
objectively satisfying the relation, that is, $m_s \gg m_{d} \sim m_u$, the apparently larger
factorizable-emission contributions induced by the vector decay constants $f_{\kappa^+},
f_{K_0^*(1430)^+} \propto (m_s - m_u)$~\cite{Cheng:2005nb} are therefore produced in
$B_c^+ \to \eta_c (\kappa, K_0^*(1430))^+$ channels, relative to the factorizable-emission
amplitudes in $B_c^+ \to \eta_c a_0(980, 1450)^+$ modes proportional to the value of
$(m_d- m_u)$, namely, tiny broken isospin symmetry.
The above iPQCD results predicted for the $B_c^+ \to \eta_c S^+$ decays await the future experimental
tests, which could help us to further explore the complicated QCD dynamics potentially.

Lastly, we come to the discussions of $B_c$ decays to $\eta_c$ plus a light tensor state.
For the $B_c^+ \to \eta_c T^+$ modes, to our best understanding, tensor states cannot be produced
via the vector current and
the conventionally factorizable-emission contributions at leading order are thus
vanished naturally. Therefore, their studies must go beyond naive factorization.
Then, the $B_c^+ \to \eta_c T^+$ BRs calculated in the iPQCD formalism are
\beq
{\rm BR}(B_c^+ \to \eta_c a_2(1320)^+)&=&
1.33
^{+0.41}_{-0.32}(\beta_{B_c})
^{+0.16}_{-0.16}(f_M)
^{+0.05}_{-0.05}(V_{cb})
\times 10^{-4} \;,
\\
{\rm BR}(B_c^+ \to \eta_c {K_2^*}(1430)^+)&=&
8.21^{+2.56}_{-2.00}(\beta_{B_c})
^{+0.79}_{-0.76}(f_M)
^{+0.33}_{-0.29}(V_{cb})
\times 10^{-6} \;,
\eeq
associated with the ratios
\beq
\frac{{\rm BR}(B_c^+ \to \eta_c a_2(1320)^+)}{{\rm BR}(B_c^+ \to J/\psi \pi^+)}
&=&
0.11^{+0.05}_{-0.04} \;, \qquad
R^{\eta_c }_{a_2/\pi} \equiv
\frac{{\rm BR}(B_c^+ \to \eta_c a_2(1320)^+)}{{\rm BR}(B_c^+ \to \eta_c \pi^+)}=
0.07^{+0.00}_{-0.01} \;,
\eeq
where the dominant errors come from the $B_c$-meson distribution amplitude shape parameter $\beta_{B_c}$.
Future tests of these values, which are solely from nonfactorizable-emission contributions, would help us examine the reliability of this iPQCD formalism.

Similar to $R_{K^*/\rho}^{\eta_c}$ in the $B_c^+ \to \eta_c V^+$ sector, another ratio
$R^{\eta_c}_{K_2/a_2}$ in the $B_c^+ \to \eta_c T^+$ decays is also defined by utilizing
the $B_c^+ \to \eta_c a_2(1320)^+$ and $B_c^+ \to \eta_c {K_2^*}(1430)^+$ BRs in the iPQCD framework,
and its value is then read as
\beq
R^{\eta_c}_{K_2/a_2}&\equiv&
\frac{{\rm BR}(B_c^+ \to \eta_c {K_2^*}(1430)^+)}{{\rm BR}(B_c^+ \to \eta_c a_2(1320)^+)}=
(6.17^{+0.20}_{-0.13}
) \times 10^{-2} \;.
\eeq
This result is induced by only nonfactorizable-emission contributions and is very close to the value
presented in Eq.~(\ref{eq:rrhkst-th}), as well as that naively anticipated in factorization ansatz
for factorizable-emission predominated $B_c^+ \to \eta_c (\rho, K^*)^+$ modes within errors.

Analogous to $B_c^+ \to \eta_c b_1(1235)^+$ but different from $B_c^+ \to \eta_c a_0(980, 1450)^+$,
the considerably constructive
interferences due to the antisymmetric leading-twist distribution amplitude between the two
nonfactorizable-emission diagrams in Figs.~\ref{fig:fig1}(c) and \ref{fig:fig1}(d) work in the
$B_c^+ \to \eta_c a_2(1320)^+$ mode indeed.
The future measurements on the iPQCD predictions of $B_c^+ \to \eta_c L^+$ in this work could
help test, even differentiate, the reliability of the adopted approaches in the literature,
which might help us to further understand the rich QCD dynamics in $B_c$-meson decays
through pulling together disparate ideas.


\section{Conclusions} \label{sec:summary}

In summary, we have studied the $B_c$-meson decays into $\eta_c$ plus a light charged meson
in the self-consistent iPQCD formalism at leading order of strong coupling $\alpha_s$. The
numerical results and phenomenological insights on {\it CP}-averaged BRs in association
with relative ratios are presented explicitly. The $B_c^+ \to \eta_c (\pi, \rho)^+$
and $B_c^+ \to \eta_c a_1(1260)^+$ BRs are generally
larger than those $B_c \to J/\psi$ decay ones correspondingly in the iPQCD formalism, which
reveal the remarkably different QCD dynamics between these two kinds of $B_c$-meson decays.
For (near-)future tests at experiments, the BRs of multibody modes of $B_c^+ \to \eta_c (\pi,
\rho)^+$ via $\eta_c$ resonance through $\eta_c \to p\bar p$ and $\eta_c \to \pi^+\pi^-
(\pi^+ \pi^-, K^+ K^-, p\bar p)$ are also predicted under the narrow-width approximation.
The experimental search performed with the successfully upgraded LHCb detector for the BRs around
${\cal O}(10^{-6})$ and above, for example,
${\rm BR}(B_c^+ \to \eta_c (\to p\bar p) \pi^+) = (2.70^{+0.74}_{-0.59}) \times 10^{-6}$,
${\rm BR}(B_c^+ \to \eta_c (\to p\bar p) \rho^+) = (7.14^{+2.03}_{-1.60}) \times 10^{-6}$,
${\rm BR}(B_c^+ \to \eta_c [\to 2(\pi^+\pi^-)] \pi^+) = (1.95^{+0.59}_{-0.50}) \times 10^{-5}$,
${\rm BR}(B_c^+ \to \eta_c [\to 2(\pi^+\pi^-)] \rho^+) = (5.16^{+1.62}_{-1.34}) \times 10^{-5}$,
$\cdots$, are expected to help probe the relevant decay channels and explore the nature of
charmonium $\eta_c$.  The almost equal BRs
of $B_c^+ \to (J/\psi, \eta_c) b_1(1235)^+$, while the surprisingly smaller ones of $B_c^+ \to
\eta_c a_0(980, 1450)^+$ compared to those of $B_c^+ \to J/\psi a_0(980, 1450)^+$ need further
explorations in both aspects of theory and experiment. Moreover, the predicted ratios $R_{K/\pi}^{\eta_c}$
and $R_{K^*/\rho}^{\eta_c}$ matching the anticipations based on the naive ansatz means that the $B_c^+ \to
\eta_c (P, V)^+$ decays are predominated by factorizable-emission contributions. The experimental
tests on the iPQCD ratios $R_{\eta_c/J/\psi}^{\pi} = 1.74^{+0.66}_{-0.50}$ and ${\rm BR}(B_c^+
\to \eta_c S^+)_{\Delta S=0}/{\rm BR}(B_c^+ \to \eta_c S^+)_{\Delta S=1}
\sim {\cal O}(0.01)$ might be urgently demanded, which could help decipher the QCD dynamics
involved in the considered $B_c$-meson decays as well as in the related charmonia $\eta_c$
and $J/\psi$ greatly.


\begin{acknowledgments}

X.L. thanks Vanya~Belyaev and Pei-Lian~Li for their helpful communications.
This work is supported by the National Natural Science
Foundation of China under Grant No.~11875033 and
by the Qing Lan Project of Jiangsu Province (Grant No.~9212218405).

\end{acknowledgments}





\end{CJK*}
\end{document}